\documentclass[lettersize,journal]{IEEEtran}
\IEEEoverridecommandlockouts

\usepackage{cite}
\usepackage{amsmath,amssymb,amsfonts}
\usepackage{algorithmic}
\usepackage{graphicx}
\usepackage{textcomp}

\usepackage{xcolor}

\newcommand{\abs}[1]{\left| #1 \right|}
\usepackage{hyperref}
\usepackage{placeins}
\usepackage[ subrefformat=parens,labelformat=parens, caption=false, font=footnotesize]{subfig}
\usepackage{mathtools}
\usepackage{units}
\usepackage[cmintegrals]{newtxmath}

\def\BibTeX{{\rm B\kern-.05em{\sc i\kern-.025em b}\kern-.08em
    T\kern-.1667em\lower.7ex\hbox{E}\kern-.125emX}}
\makeatother

\makeatletter
\def\ps@IEEEtitlepagestyle{%
  \def\@oddfoot{\mycopyrightnotice}%
  \def\@oddhead{\hbox{}\@IEEEheaderstyle\leftmark\hfil\thepage}\relax
  \def\@evenhead{\@IEEEheaderstyle\thepage\hfil\leftmark\hbox{}}\relax
  \def\@evenfoot{}%
}
\makeatother

\def\mycopyrightnotice{%
  \begin{minipage}{\textwidth}
  \centering \scriptsize
    This work has been accepted by the IEEE Communications Letters for publication.  Copyright may be transferred without notice, after which this version may no longer be accessible.
  \end{minipage}
}

\begin{document}

\title{Performance Analysis of Data Detection in the THz-Band under Channel-Correlated Noise}
 
\author{
        Almutasem Bellah Enad, \IEEEmembership{Graduate Student Member, IEEE,}
        Hadi Sarieddeen, \IEEEmembership{Senior Member, IEEE,}
        Jihad Fahs, \IEEEmembership{Member, IEEE,}
        Hakim Jemaa, \IEEEmembership{Graduate Student Member, IEEE}
        and Tareq Y. Al-Naffouri, \IEEEmembership{Fellow Member, IEEE}%
        
\thanks{A.~B.~Enad, H.~Sarieddeen, and J.~Fahs, are with American University of Beirut (AUB): aae118@mail.aub.edu, \{hadi.sarieddeen, jihad.fahs\}@aub.edu.lb.
H. Jemaa and T. Y. Al-Naffouri are with King Abdullah University of Science and Technology (KAUST): \{hakim.jemaa, tareq.alnaffouri\}@kaust.edu.sa.
This work is supported by the AUB's University Research Board and Vertically Integrated Projects Program, and KAUST's Office of Sponsored Research under Award No.~ORFS-CRG12-2024-6478.}}

\maketitle

\begin{abstract}
We present a comprehensive symbol error rate (SER) analysis framework for link-level terahertz (THz)-band communication systems under linear zero-forcing (ZF) data detection. First, we derive the mismatched SER for indoor THz systems under independent channel and noise assumptions, calculating the probability density function of the ratio of Gaussian noise to $\alpha$-$\mu$ channels resulting from ZF filtering. Next, we derive the precise SER under correlated channel and noise conditions, modeling dependencies using the copula method. Finally, we evaluate the SER for THz channels with correlated distortion noise from hardware impairments. Simulations demonstrate that the proposed framework corrects for multi-dB SERs resulting from the channel-noise independence assumption.
\end{abstract}


\section{introduction}
\IEEEPARstart{T}{erahertz} (THz) communications will play a central role in next-generation wireless systems. Spanning 0.3 to 3 THz, the THz band promises terabit-per-second (Tbps) data rates, supporting diverse communication and sensing applications~\cite{Jornet2024Evolution,Sarieddeen2021Overview}. However, achieving ultra-reliability, low latency, and high energy efficiency remains challenging \cite{Sarieddeen2024Bridging}, and performance analysis frameworks are still not mature.

At the link level, linear data detection methods are favored for their computational efficiency, essential under stringent Tbps constraints. Yet, their performance under THz-specific channels and noise remains underexplored. Zero-forcing (ZF) detection, which reduces interference by applying the pseudo-inverse of the channel, generally yields higher symbol error rates (SERs) than minimum mean squared error (MMSE) detection, particularly at low signal-to-noise ratios (SNRs)~\cite{proakis2008digital}. However, ZF offers lower complexity and does not require SNR knowledge. Although most linear detection SER studies assume Gaussian noise post-filtering, optimizing THz systems demands accurate channel and noise modeling~\cite{8610080,papasotiriou2021experimentally}. 

In practical THz systems, channel-noise correlation can arise from hardware imperfections causing transmit and receive noise distortions or from channel-induced molecular absorption noise~\cite{8610080,Sarieddeen2021Overview}. Traditional independent noise models fail to capture such effects~\cite{Jemaa2024,Magableh2009}. Incorporating channel-correlated noise is more physically consistent and increasingly relevant in multi-antenna deployments, addressing a key gap in the literature. Small-scale fading in indoor THz channels can be effectively modeled by the $\alpha$-$\mu$ distribution with flexible parameters~\cite{papasotiriou2021experimentally,1048550}, which has been used in frameworks incorporating misalignment fading~\cite{8610080,10018285} and random fog effects~\cite{9714471}. Copula theory~\cite{bickel2009copula,nelsen2006introduction} has been widely adopted in wireless communications to model complex dependencies, including Nakagami-$m$ fading with tail dependence~\cite{4487493}, and non-standard inter-band fading~\cite{7032317}. Copula-based methods have also been used to model correlation in the context of multiple access channels~\cite{9762969}, intelligent-surface-aided systems~\cite{9690184}, and physical layer security in wiretap channels~\cite{9159617}.

In this paper, we propose a SER analysis framework for single-input single-output (SISO) THz links, addressing both correlated and independent channel and noise scenarios. While SER analysis is well-established in the literature~\cite{Jemaa2024,9714471,8610080}, our work extends it by incorporating THz-specific challenges such as $\alpha$-$\mu$ fading, hardware impairments, and channel-correlated noise. These factors are often overlooked in conventional analyses but are essential for a realistic assessment of THz system performance. Our contributions are threefold. First, we analyze the SER under independent channel and noise assumptions by deriving the probability density function (PDF) of the ratio between Gaussian noise and the $\alpha$-$\mu$ channel. Second, we evaluate the SER under correlated channel and noise conditions using the copula method, which captures statistical dependencies for more realistic error analysis. Finally, we adopt a practical THz communication model, derive the corresponding SERs accounting for channel–distortion noise correlation due to hardware impairments, and complement the analysis with asymptotic results.
Throughout the paper, $\abs{\cdot}$ represents the absolute value, $\mathbb{E}[\cdot]$ is the expectation operator, and $\mathcal{CN}(0,2 \sigma^2)$ denotes circularly symmetric complex Gaussian distributions with independent and identically distributed real and imaginary components, each with variance $\sigma^2$. $Q(x)\!=\!\frac{1}{\sqrt{2 \pi}} \int_x^{\infty} e^{ \left(-\frac{u^2}{2}\right)} d u$ is the $Q(.)$ function \cite{prudnikov1986integrals}, $\Gamma(.)$ is the gamma function, and $\gamma(\zeta, u)\!=\!\int_0^u t^{\zeta-1} e^{-t} d t$ is the lower incomplete gamma function. $G_{p, q}^{m, n}\left[z \left\lvert \begin{smallmatrix}a_1, \ldots, a_p \\ b_1, \ldots, b_q\end{smallmatrix} \right.\right]$ is the Meijer G-function \cite[eq. (9.301)]{prudnikov1986integrals}.

\section{system Model and Problem Formulation}
We analyze the impact of key THz-specific factors on system performance, providing insights into the robustness of THz communications. Specifically, we compare two models: a standard model capturing fast fading via the $\alpha$-$\mu$ distribution~\cite{8610080,9714471}, and a specialized model accounting for hardware impairments as distortion noise~\cite{8610080}. In both cases, $\alpha$-$\mu$ fading effectively models small-scale THz channel variations.

\subsection{Generic THz system model}

We consider a system of baseband input-output relation,
\begin{equation}
    r=\sqrt{p_tG_tG_r} h_p h_f s + n = \nu h_f s + n = h s + n, \label{sys1}
\end{equation}
where $r$ is the received symbol, $p_t$ is the average transmit power, $G_t$/$G_r$ are the transmit/receive antenna gains, and $s$ is the transmitted quadrature amplitude modulation (QAM) symbol. The additive noise, $n \sim \mathcal{CN}(0,2 \sigma^2)$, has a Rayleigh-distributed amplitude with a PDF 
\begin{align}
    f_{|n|}(x) &= \frac{x}{\sigma^2} \exp\left(-\frac{x^2}{2\sigma^2}\right), \quad \text{for}\ \, x \geq 0.\label{PDF noise}
\end{align}
The THz channel is characterized by $h_p$ and $h_f$. Here, $h_p$ represents free-space path loss due to spreading and molecular absorption, given by \(h_p = \left(c/(4 \pi f d)\right)^{\frac{\varrho}{2}} e^{-\frac{1}{2} K_{\text{abs}} d},\) where $c$ is the speed of light, $f$ is the operating frequency, $d$ is the communication distance, and $K_{\text{abs}}$ is the molecular absorption coefficient (more details in \cite{9591285}). In measurement-based sub-THz/THz works \cite{9591285,Jemaa2024}, the path loss exponent, $\varrho$, is best-fit to $2$. Finally, $h_f$ represents complex small-scale fading of $\alpha$-$\mu$-distributed magnitude \cite{Magableh2009}. Defining $\nu\!\triangleq\!\sqrt{p_tG_tG_r} h_p$, the PDF and CDF of $h\!=\!\nu h_f$ can be expressed as
\begin{align}
    f_{|h|}(y)&=\!\frac{\alpha \mu^\mu y^{\alpha \mu - 1}}{(\hat{Z}\nu)^{\alpha \mu} \Gamma(\mu)}  \exp\left(-\mu y^\alpha/(\nu\hat{Z})^{\alpha}\right),\label{PDF alpha-mu}\\
    F_{|h|}(y)&=\frac{\gamma{(\mu,\mu y^\alpha / (\hat{Z}\nu)^\alpha)}}{\Gamma(\mu)},\label{CDF alpha-mu}
\end{align}
where $\alpha\!>\! 0$ is a fading parameter, $\mu$ is the normalized variance of the fading channel, and $\hat{Z} = \sqrt[\alpha]{\mathbb{E}[|h_f|^\alpha]}$ is the $\alpha$ root mean value of the fading channel.

\subsection{Specific THz system model}
We consider a model that incorporates THz distortion noise from hardware impairments, capturing the dependency between channel fading and receiver noise~ \cite{8610080,10891806},
\begin{equation}
    r = \sqrt{p_tG_tG_r} h_p h_f (s + n_t) + n_r + n= h(s + n_t) + n_r + n, \label{sys2}
\end{equation}
where $\kappa_t, \kappa_r >0$ represent hardware imperfections, $n_t \!\sim\! \mathcal{CN}(0, \kappa_t^2)$ and $ n_r \!\sim\! \mathcal{CN}(0, \kappa_r^2|h|^2)$ are distortion noise components from transmit and receive impairments, respectively, and $n \!\sim\! \mathcal{CN}(0, 2 \sigma^2)$ is independent of $h$. Here, $\abs{h}$ also follows an $\alpha$-$\mu$ distribution, but only $n_r$ is correlated with $h$.

\subsection{Problem formulation}
At the receiver, a linear ZF detector equalizes (inverts) the channel. The post-filtering generic system model~\eqref{sys1} is thus
\begin{equation}
    \tilde{r} = h^{-1}r  = s+n/h  = s + z, \label{z1}
\end{equation}
where $z \!=\! n/h$ is the resultant effective noise. Similarly, the post-filtering system model in \eqref{sys2} is
\begin{equation}
    \Tilde{r}= h^{-1}r = s + \frac{h n_t + n_r + n}{h}= s + n_t + n_r' + \frac{n}{h}= s +z, \label{z2}
\end{equation}
where \(n_r'\!=\!n_r/h \!\sim\! \mathcal{CN}(0, \kappa_r^2)\). Our aim is to develop a generic framework for calculating the SER for these system models without necessarily assuming a channel-independent noise. We distinguish three scenarios for which we derive the PDFs of the corresponding effective noise variables and leverage them in finding the SERs.

\section{Proposed SER Analyses}

\subsection{Independent channel and noise model}\label{sec21}
We compute the SER by analyzing the statistics of the ratio between the additive noise and the channel. For ZF detection in indoor THz channels, the effective noise variable in~\eqref{z1} is modeled as the ratio of Gaussian noise to the $\alpha$-$\mu$ fading,
\begin{equation}
    z=\frac{|n|}{|h|} \exp\left(j (\theta_n-\theta_h)\right)  =\uprho \exp\left(j \theta_z\right) =z_r+iz_i \label{ratio sys1},
\end{equation}
where $n \!=\! |n|  \exp\left(j \theta_n\right) \!\sim\! \mathcal{CN}(0,2\sigma^2)$ and $h \!=\! |h|  \exp\left(j \theta_h\right) $ is channel fading of a magnitude distribution PDF defined in~\eqref{PDF alpha-mu}. Under statistically independent channel and noise, the phase term, $\theta_z$, follows a uniform distribution over $[0, 2\pi]$, as both $\theta_n$ and $\theta_h$ are uniformly distributed~\cite{pewsey2013circular}. Since \(\rho\) and \(\theta_z\) are independent random variables, their joint PDF is given by
\begin{align}
    &f_{\rho,\theta_z}(\rho,\theta)\notag\\
    &= \frac{1}{2\pi}\!\!\int_{0}^{+\infty}\!\!\!\!y f_{|n|}(\rho y) f_{|h|}(y) dy\label{integral-ind}\\
    &\stackrel{(\mathrm{a})}{=} \frac{\alpha \mu^\mu \rho}{2\pi\sigma^2(\nu\hat{Z})^{\alpha\mu}\Gamma(\mu)}\!\int_{0}^{+\infty}\!\!\!y^{\alpha\mu+1}\!\exp\!\left(-\frac{\rho^2y^2}{2\sigma^2}\!-\!\frac{\mu y^\alpha}{(\nu\hat{Z})^{\alpha}}\right)dy\notag\\
    &\stackrel{(\mathrm{b})}{=} \frac{\ \mu^\mu \rho}{2\pi\sigma^2\Gamma(\mu)\cdot (\nu\hat{Z})^{\alpha \mu}}  \int_{0}^{+\infty}\!\!\!t^{\frac{2}{\alpha}+\mu-1} \exp\!\left(-\frac{\rho^2t^{\frac{2}{\alpha}}}{2\sigma^2}\!-\!\frac{\mu t}{(\nu\hat{Z})^{\alpha}}\right)dt\notag\\
    &\stackrel{(\mathrm{c})}{=}\frac{\alpha \mu^\mu \rho}{2\pi\sigma^2\Gamma(\mu)(\nu\hat{Z})^{\alpha \mu}}  \int_{0}^{+\infty} \!\!\!t^{\frac{2}{\alpha}+\mu-1} G_{0,1}^{1,0} \left( \frac{\rho^2}{2\sigma^2}t^{\frac{2}{\alpha}} \middle| \begin{array}{c} \sim \\ 0 \end{array} \right)\notag\\&\hspace{+40mm} \times G_{0,1}^{1,0} \left( \frac{\mu}{(\nu\hat{Z})^{\alpha}} t \middle| \begin{array}{c} \sim \\ 0 \end{array} \right) \, dt\notag\\
    & \stackrel{(\mathrm{d})}{=}\eta\, \rho \, G_{l,k}^{k,l} \,\left( \zeta \rho^{2k} \middle| \begin{array}{c} I(l,1-\mu-\frac{2}{\alpha}) \\ I(k,0) \end{array} \right), \label{PDF_rho_theta}
\end{align}
where 
\begin{equation}
\eta = \frac{(\nu\Hat{Z})^{2} \cdot k^{\frac{1}{2}} \cdot l^{\frac{2}{\alpha} + \mu - \frac{1}{2}}}{(2\pi)^{\frac{l+k}{2}} \sigma^2 \Gamma(\mu) \mu^{\frac{2}{\alpha}}}, \quad
\zeta = \frac{{\left((\nu\hat{Z})^\alpha l\right)}^l}{(2k\sigma^2)^k \cdot \mu^l}\notag.
\end{equation}
In \(\stackrel{(\mathrm{a})}{=}\), we substitute equations \eqref{PDF noise} and \eqref{PDF alpha-mu} into \eqref{integral-ind}. In \(\stackrel{(\mathrm{b})}{=}\), we use a change of variable \(t \!=\! y^\alpha\). In \(\stackrel{(\mathrm{c})}{=}\), the exponential function is rewritten using its Meijer \(G\)-function representation~\cite{prudnikov1986integrals}, as done in\cite{ELSAYED2024130558,Jemaa2024}, \(e^{-x} = G_{0,1}^{1,0} \left[x \Bigg| \begin{array}{c} \sim \\ 0 \end{array} \right]\). Finally, in \(\stackrel{(\mathrm{d})}{=}\), we apply the Meijer \(G\)-function integration property~\cite{wolfram}. Here, \(I(n, q) = \frac{q}{n}, \frac{q+1}{n}, \ldots, \frac{q + n - 1}{n}\), and \(\frac{2}{\alpha} = \frac{l}{k}\), where \(l\) and \(k\) are coprime (ensuring non-integer values of \(\alpha\) are accounted for). The parameter \(\mu\) can assume both integer and non-integer values.
Furthermore, we express the joint PDF of the real and imaginary components, $z_r$ and $z_i$, as
\begin{align}
&f_{\Re(z), \Im(z)}(z_r, z_i) = \frac{1}{\rho}f_{\rho,\theta_z}(\rho,\theta)\notag \\ &=\eta\cdot G_{l,k}^{k,l} \left( \zeta (z_r^2+z_i^2)^{k} \middle| \begin{array}{c} I(l,1-\mu-\frac{2}{\alpha}) \\ I(k,0) \end{array} \right).\label{PDF ratio sys1}
\end{align}
Without loss of generality, we assume the first symbol, $s_1$, of a quadrature phase shift keying (QPSK) constellation is transmitted (all symbols are equiprobable and yield the same error rate due to symmetry). Let $P_{c|s_1}$ and $P_{e|s_1}$ denote the probabilities of correctly and erroneously detecting $s_1$. Then,
\begin{equation}
    P_{e|s_1}= 1 - P_{c|s_1}= 1-\iint_{D_1} f_{\Re(z), \Im(z)}(r_x-\delta, r_y - \beta)\, dr_y dr_x,\label{PE}
\end{equation}
where $r_x$ and $r_y$ denote the real and imaginary components of the received signal, $r$, and $ D_1 = [0, \infty) \times [0, \infty) $ corresponds to the constellation region of symbol $ s_1 = \delta + i\beta $, for $\delta, \beta > 0$. Finally, using~\eqref{PDF ratio sys1} and~\eqref{PE}, we get
\begin{align}
&P_{c|s_1}\!=\!\eta\!\!\iint_{D_1}\!\!\!\!\! G_{l,k}^{k,l}\left(\!\zeta(\!(r_x\!-\!\delta)^2\!\!+\!\!(r_y\!-\!\beta)^2)^{k} \middle|\!\!\begin{array}{c} I(l,1\!-\!\mu\!-\!\frac{2}{\alpha}) \\ I(k,0) \end{array} \!\!\!\right) \!dr_y dr_x.
    \label{SER 1st}
\end{align}

\subsection{Dependent channel and noise model}
\label{sec:dependent}
We model the correlation statistically using the copula method \cite{bickel2009copula,nelsen2006introduction}. The joint PDF of $n$ and $h$ is expressed as
\( f_{n,h}(x,y) \! =\! C(F_n(x), F_h(y)) f_n(x) f_h(y),\)
where $f_h(y)$, $f_n(x)$ and  $F_h(y)$, $F_n(x)$ are marginal PDFs and CDFs of $h$ and $n$, respectively, and \( C(\cdot,\cdot) \) denotes the bivariate copula density function. There are multiple families of copulas; this paper focuses on both the  FGM and Frank methods. An FGM copula models weak dependence between two variables as \cite{nelsen2006introduction}
\begin{equation}
    C(u, v) = 1 + \vartheta (2u - 1)(2v - 1),
    \label{eq:FGM}
\end{equation}
where \(\vartheta \in [-1, 1]\) controls the strength and direction of dependence. The Frank copula models stronger dependencies,
\begin{equation}
    C(u, v)\!=\!-\lambda\!\left(e^{-\lambda}\!-1\right)\!\left( \frac{e^{-\lambda(u + v)}}{\left( \left( e^{-\lambda u} - 1 \right) \left( e^{-\lambda v} - 1 \right) + \left( e^{-\lambda} - 1 \right)\!\right)^2 } \right),
    \label{eq:frank}
\end{equation}
where \(\lambda \!\in\! (-\infty, \infty) \!\setminus\! \{0\}\) governs dependence strength, with positive values indicating positive dependence. We use the copula method to model the correlation between the \(\alpha\)-\(\mu\) ($|h|$) and Rayleigh ($|n|$) distributions. Let \(\xi\! \overset{\triangle}{=}\!\frac{\alpha \mu^\mu}{(\nu\hat{Z})^{\alpha \mu} \Gamma(\mu)}\), then
\begin{align}
    &f_{|n|,|h|}(x,y)\!=\!\frac{\xi}{\sigma^2} x\!y^{\alpha\mu-1}\!\exp\!\left(\!-\frac{x^2}{2\sigma^2}\!-\!\!\frac{\mu\,y^\alpha}{(\nu\!\hat{Z})^{\alpha}}\!\right)\!C(F_n(x), F_h(y)). \notag
\end{align}
Similar to the derivation of equations~\eqref{PDF_rho_theta} and~\eqref{PDF ratio sys1}, we find 
\begin{align}
    f_{\Re(z), \Im(z)}(z_r, z_i)&=\frac{\xi}{2\pi \sigma^2}\hspace{-0.7pt}\!\int_0^\infty\!\hspace{-3pt}y^{\alpha\mu +1}\!\exp\!\left(\!-\frac{{(z_r^2\!+\!z_i^2)\!y^2}}{2\sigma^2}\!-\!\frac{\mu\,y^\alpha}{{(\nu\hat{Z})}^{\alpha}}\right)\notag\\
    &\times
    C(F_{|n|}(\sqrt{z_r^2+z_i^2}\, y), F_{|h|}(y)) dy, \label{eq:sim}
\end{align}
which gives $P_e$ using~\eqref{PE}. The copula term, $C(F_{|n|}(\sqrt{z_r^2+z_i^2}\, y), F_{|h|}(y))$, is evaluated according to~\eqref{eq:FGM} and~\eqref{eq:frank} for the FGM and Frank methods, respectively.

\subsection{A specific THz model}\label{sec23}

We next consider the system model in~\eqref{sys2}, which includes distortion noise from THz hardware impairments exhibiting receiver-side channel dependency~\cite{8610080}. This model can also be extended to include molecular absorption noise~\cite{Sarieddeen2021Overview}. The resulting post-filtering effective noise in~\eqref{z2} simplifies to $z\!=\!n_t \!+\! n_r' \!+\! \frac{n}{h}\!=\!\varphi\!+\!\frac{n}{h}\!=\!\varphi\!+\!\omega$, where $\varphi \!\sim\! CN(0, 2 \sigma_t^2 )$, \(2\sigma_t^2 \!=\!(\kappa_t^2 \!+\! \kappa_r^2)\), and 
\begin{equation}
f_\varphi(\varphi_r,\varphi_i) = \frac{1}{2\pi \sigma_t^2} \exp\left(-\frac{\varphi_r^2 + \varphi_i^2}{2\sigma_t^2}\right).
\end{equation}
Also, we express $\omega=\frac{n}{h}$ via the joint PDF~\eqref{PDF_rho_theta},
\begin{align}
f_{\uprho,\theta_\omega}(\rho,\theta) &=\frac{\xi\rho}{2\pi\sigma^2} \int_{0}^{+\infty} y^{\alpha \mu+1}\!\exp\left( -\frac{\rho^2y^2}{2\sigma^2}-\mu \frac{y^\alpha}{(\nu\hat{Z})^{\alpha}}\right)\,dy,
\end{align}
where \(\xi\) is given in Section~\ref{sec:dependent}. In Cartesian coordinates,
\begin{align}
f_{\omega}(\omega_r,\omega_i)\!=\!\frac{\xi}{2\pi\sigma^2} \!\int_{0}^{+\infty}\!\!y^{\alpha \mu+1}\!\exp\!\left(\!-\frac{(\omega_r^2\!+\!\omega_i^2)\!y^2}{2\sigma^2}\!-\!\mu\!\frac{y^\alpha}{(\nu\!\hat{Z})^{\alpha}}\!\right)\!dy.
\end{align}
It follows that the PDF of $z=\varphi+w$ is the convolution between the PDFs of $\varphi$ and $\omega$. Therefore,
\begin{align}
    &f_{\Re(z), \Im(z)}(z_r, z_i) = ( f_{\varphi} * f_w )(z_r,z_i) \notag\\ &=\iint_{-\infty}^{\infty}f_{\varphi}(\varphi_i,\varphi_r) f_w(z_r - \varphi_r,z_i - \varphi_i) \, d\varphi_i d\varphi_r \notag\\
    &=\iint_{-\infty}^{\infty} \int_{0}^{\infty}\frac{\xi}{2\pi \sigma_t^2\cdot2\pi\sigma^2}y^{\alpha \mu+1}\exp{\left(-\frac{\mu \,y^\alpha}{(\nu\hat{Z})^{\alpha}}\right)}\notag\\
    &\times\exp\left(-\frac{(\varphi_r^2+\varphi_i^2)}{2\sigma_t^2} \hspace{-0.3pt}-\hspace{-0.3pt}\frac{((z_r - \varphi_r)^2 + (z_i - \varphi_i)^2)y^2}{2\sigma^2}\right) \hspace{-1.2pt} dy d\varphi_i d\varphi_r.\label{A5}
\end{align}
We simplify this integration by first solving
\begin{align}
I&=\iint_{-\infty}^{\infty}\!\exp\left(-\frac{(z_r\!-\!\varphi_r)^2\!+\!(z_i\!-\!\varphi_i)^2}{2\sigma^2}\!y^2-\frac{\varphi_r^2\!+\!\varphi_i^2}{2\sigma_t^2}\right) d\varphi_r d\varphi_i\notag\\
&= \exp\left( -\frac{y^2}{2 \sigma^2} (z_r^2 + z_i^2) \right)\times \notag\\&\iint_{-\infty}^{\infty}\hspace{-3pt}\!\exp\left(\!-\left(\frac{y^2}{2 \sigma^2}\!+\!\frac{1}{2\sigma_t^2}\!\right) (\varphi_r^2\!+\!\varphi_i^2)\!+\!\frac{y^2}{\sigma^2}\!(z_r\!\varphi_r\!+\!z_i\!\varphi_i) \right)\,d\varphi_r d\varphi_i. \notag
\end{align}
Using \cite[Eq 3.323.2]{zwillinger2014table} and substituting the result of the integration over $d\varphi_r$ and $d\varphi_i$ in~\eqref{A5} yields the PDF of $z$,
\begin{align}
    &f_{\Re(z),\Im(z)}(z_r, z_i)\nonumber\\
    &=\int_{0}^{\infty}\hspace{-3pt}\frac{\xi y^{\alpha \mu-1}}{2\pi(\sigma_t^2+\frac{\sigma^2}{y^2})} \exp\left(-\frac{(z_r^2+z_i^2)}{2(\sigma_t^2+\frac{\sigma^2}{y^2})}-\frac{\mu\,y^\alpha}{(\nu\hat{Z})^{\alpha}}\right)dy.
\end{align}
Finally, the average SER can be computed following~\eqref{PE} as
\begin{align}
    P_{e|s_1}&= 1 -\!\iint_0^\infty\!\int_{0}^{\infty}\!\frac{\xi}{2\pi (\sigma_t^2\!+\!\frac{\sigma^2}{y^2})}y^{\alpha\mu\!-\!1}\notag\\& \quad \exp \left(-\frac{((r_x-\delta)^2 + (r_y-\beta)^2)}{2(\sigma_t^2+\frac{\sigma^2}{y^2})}\!-\!\frac{\mu\,y^\alpha}{(\nu\hat{Z})^{\alpha}}\right) dy dr_x dr_y \notag\\
    &= 1 - \xi \int_{0}^{\infty} y^{\alpha \mu-1} \exp\left(- \frac{\mu\,y^\alpha}{(\nu\hat{Z})^{\alpha}}\right) \notag\\
    &\times \left(\int_0^\infty \frac{1}{\sqrt{2\pi (\sigma_t^2+\frac{\sigma^2}{y^2})}} \exp\left(-\frac{ (r_y-\beta)^2}{2(\sigma_t^2+\frac{\sigma^2}{y^2})}\right) \, dr_y\right)\notag\\
    &\times \left(\int_0^\infty \frac{1}{\sqrt{2\pi (\sigma_t^2+\frac{\sigma^2}{y^2})}} \exp\left(-\frac{(r_x-\delta)^2}{2(\sigma_t^2+\frac{\sigma^2}{y^2})}\right)\, dr_x \right)  \, dy.\notag
\end{align}
By expressing the Gaussian integrals in terms of $Q$-functions,
\begin{align}
    P_{e|s_1} &= 1 - \xi \int_{0}^{\infty} \left(1-Q \left( \frac{\beta}{\sqrt{\sigma_t^2 + \frac{\sigma^2}{y^2}}} \right)\right)\left(1-Q \left( \frac{\delta}{\sqrt{\sigma_t^2 + \frac{\sigma^2}{y^2}}} \right)\right)\notag\\
    &\times  y^{\alpha \mu-1} \exp\left(- \frac{\mu\,y^\alpha}{(\nu\hat{Z})^{\alpha}}\right) \, dy \notag
\end{align}
\begin{align}
    P_{e|s_1}&=1-\xi\!\int_{0}^{\infty}\!y^{\alpha \mu-1}{\left(1\!-\!Q \left(\!\frac{\delta}{\sqrt{\sigma_t^2 + \frac{\sigma^2}{y^2}}} \right)\right)}^2\!\exp\left(-\frac{\mu \, y^\alpha}{(\nu\hat{Z})^{\alpha}}\right)\!dy,\label{eq:biss}
\end{align}
where~\eqref{eq:biss} is valid whenever the symbols $s$ are located at the bisectors of their corresponding QPSK quadrants (i.e., $\delta = \beta$).

We further analyze the asymptotic behavior of $P_{e|s_1}$ at high SNR. Let \( \Upsilon \triangleq \frac{\delta h}{\sqrt{\sigma_t^2 h^2 + \sigma^2}} \), then, equation~\eqref{eq:biss} becomes
\begin{align}
    P_{e|s_1}
    &= \!1-\mathbb{E}_{|h|}\left[\left(1\!-\!Q \left(\Upsilon \right)\right)^2\right]= 2\mathbb{E}_{|h|}\left[Q \left(\Upsilon\right) \right]-\mathbb{E}_{|h|}\left[Q^2 \left(\Upsilon\right)\right],\notag
\end{align}
where the PDF, \( f_{|h|}(y) \), is given in~\eqref{PDF alpha-mu}. We analyze the probability of a ``deep-fade" event. At high SNR (small $\sigma$), the dominant source of error is the overall channel gain becoming small, which occurs with probability~\cite{tse2005fundamentals}
\begin{equation}
    P_{e|s_1} \hspace{-1.5pt} \approx \mathbb{P}(\Upsilon < 1) =  \mathbb{P} \left(\hspace{-1.5pt}|h|\hspace{-1.5pt} < \hspace{-1.5pt}\frac{\sigma}{\sqrt{\delta^2-\sigma_t^2}}\hspace{-1pt}\right) \hspace{-1.5pt} = \hspace{-1.5pt} F_{|h|}\left(\hspace{-1.5pt}\frac{\sigma}{\sqrt{\delta^2-\sigma_t^2}}\hspace{-1.5pt}\right), \label{Asymptoti}
\end{equation}
where $F_{|h|}(.)$ is defined in~\eqref{CDF alpha-mu}. For small $\sigma$, the CDF argument is small. As $x \!\to\! 0$, $F_{|h|}(x) \!\approx\! \frac{1}{\Gamma(\mu)} \left( \frac{\mu x^\alpha}{(\hat{Z}\nu)^\alpha} \right)^\mu$~\cite{PARIS2002323}, resulting in an asymptotic expression $P_e \!\propto\! \left( \frac{\delta^2 - \sigma_t^2}{\sigma^2} \right)^{-\frac{\alpha \mu}{2}}$, where $\frac{\delta^2 - \sigma_t^2}{\sigma^2}$ represents the ratio between the symbol minus channel-dependent noise powers and the channel-independent noise power. The asymptotic slope, $\frac{\alpha \mu}{2}$, matches that of~\cite{10003243} for the SISO case with independent channel and noise. 

\section{Simulation Results}

\begin{figure}[t!]
    \centering
    \includegraphics[width=0.43\textwidth]{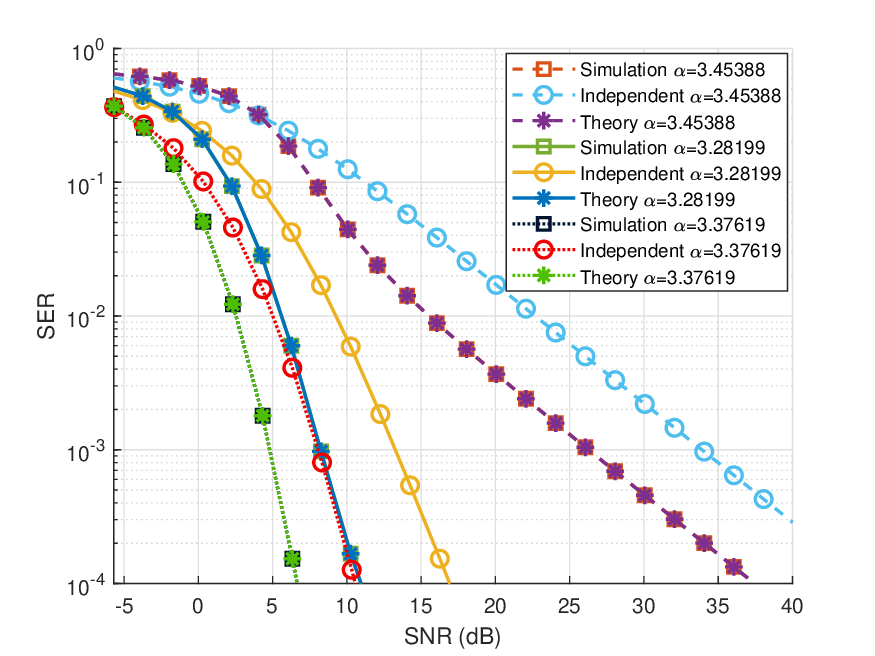}
    \caption{SERs under channel and noise correlation using the Frank method~\eqref{eq:frank}, $\lambda=7$. Theory refers to eq.~\eqref{PE} - $f_{\Re(z), \Im(z)}(\cdot,\cdot)$ given by~\eqref{eq:sim}.}
    \label{fig-frank}
\end{figure}

\begin{figure}[t!]
    \centering
    \includegraphics[width=0.43\textwidth]{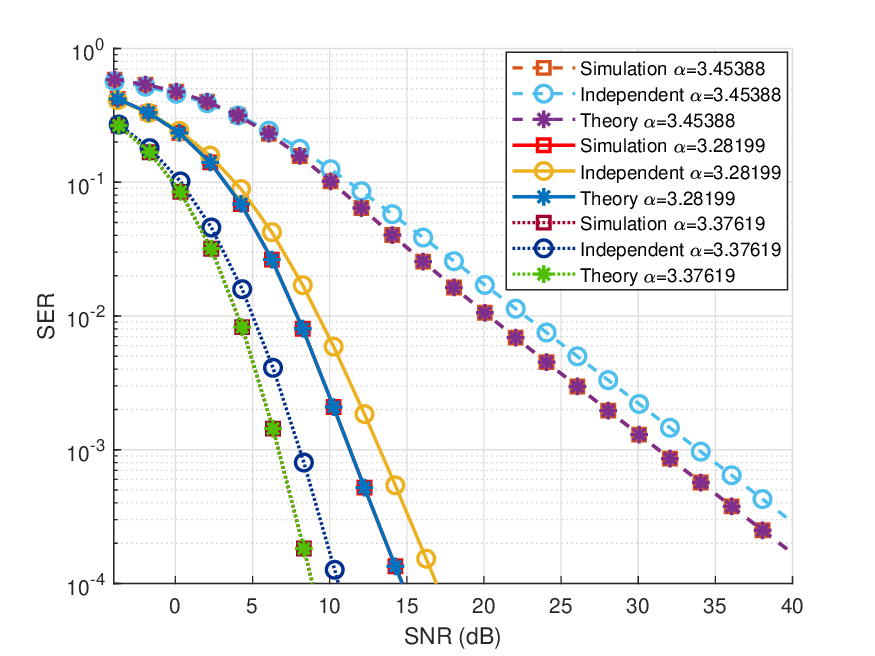}
    \caption{SERs under channel and noise correlation using the FGM method~\eqref{eq:FGM}, $\vartheta=0.9$. Theory refers to eq.~\eqref{PE} - $f_{\Re(z), \Im(z)}(\cdot,\cdot)$ is given by~\eqref{eq:sim}.}
    \label{fig-FGM}
\end{figure}

\begin{figure}[t!]
    \centering
    \includegraphics[width=0.43\textwidth]{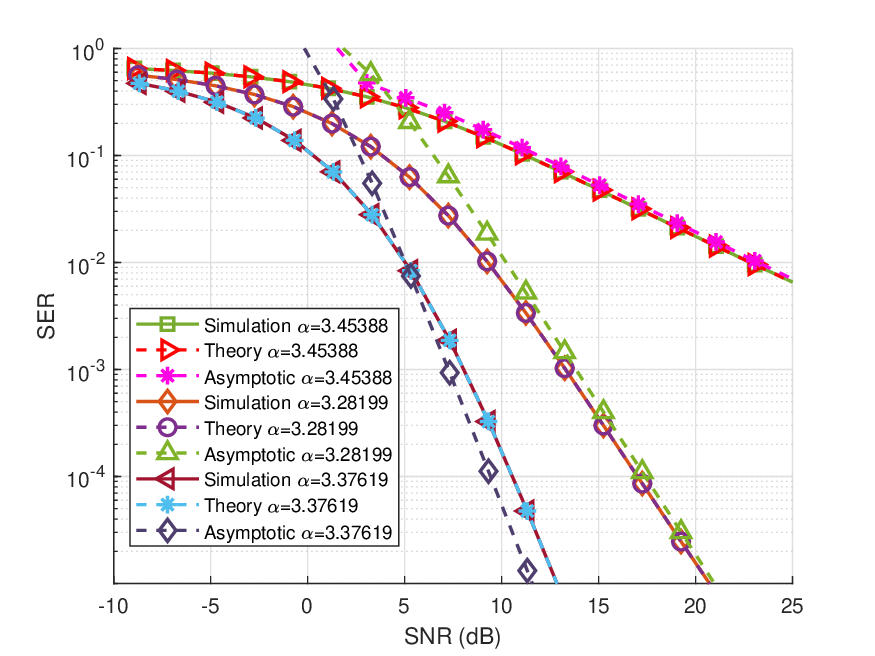}
    \caption{SERs under correlated channel and distortion noise in receivers. Theory and Asymptotic refers to eq.~\eqref{eq:biss},\eqref{Asymptoti} - $\kappa_t=\kappa_r=0.2$.}
    \label{fig-newmodel}
\end{figure}

\begin{figure}[t!]
    \centering
    \includegraphics[width=0.43\textwidth]{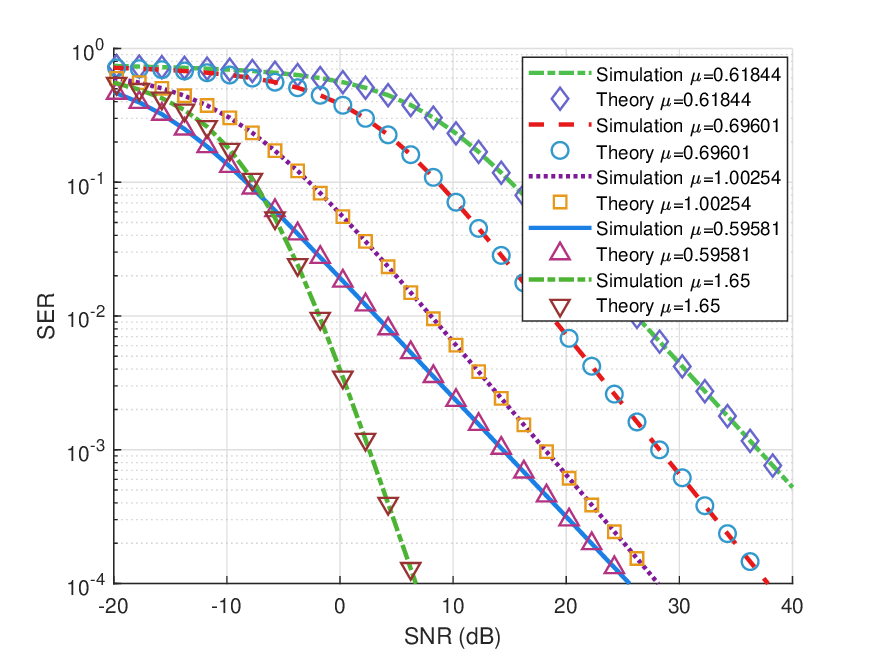}
    \caption{SERs under independent channel and noise. Theory refers to eq.~\eqref{SER 1st} - the $\alpha$ values are rounded to the closest integer.}
    \label{fig-indep}
\end{figure}

We simulate ZF detection in a THz system with fast fading modeled by the $\alpha$-$\mu$ distribution. Monte Carlo simulations with $N = 10^6$ trials are conducted. Noise power is $\sigma^2 = k_{\mathrm{B}}TB$, where $k_{\mathrm{B}} = \unit[1.38 \times 10^{-23}]{J/K}$ is Boltzmann’s constant, $T = \unit[300]{K}$ the ambient temperature, and $B = \unit[4]{GHz}$ the bandwidth, reflecting realistic thermal noise~\cite{Jemaa2024}. QPSK modulation is used with a carrier frequency $f = \unit[0.142]{THz}$~\cite{Jemaa2024,papasotiriou2021experimentally}. Distances vary based on the $ \alpha$ and $\mu$ values taken from Table~\ref{Tab1}, extracted from~\cite{papasotiriou2021experimentally,papasotiriou2021fading}, to reflect diverse propagation conditions. Antenna gains are $G_r = \unit[19]{dBi}$ and $G_t = \unit[0]{dBi}$~\cite{papasotiriou2023outdoor,Jemaa2024}, and transmit power $p_t$ is set to achieve the desired SNR range. Hardware impairments are modeled with $\kappa_t = \kappa_r = 0.2$~\cite{8610080}.
Figure~\ref{fig-frank} shows the SER for independent and correlated Frank copula cases. The theoretical curve for the Frank model matches simulations closely, while the independent assumption significantly overestimates errors, especially as $\lambda$ increases, highlighting the importance of modeling channel-noise correlation. A similar trend appears in Fig.~\ref{fig-FGM} for the FGM copula, though the deviation is smaller due to its weaker dependence, even at $\vartheta = 0.9$. In Fig.~\ref{fig-newmodel}, which accounts for THz transceiver impairments, both the SER from~\eqref{eq:biss} and its asymptotic form~\eqref{Asymptoti} closely match simulation results, validating the proposed framework.
Figure~\ref{fig-indep} further shows the SER performance under independent channel and noise, with the theoretical expression given by equation~\eqref{SER 1st}. It is important to note that the $\alpha$ values in Table~\ref{Tab1} can not be used with high precision in simulations to ensure the convergence of the integral in equation~\eqref{SER 1st}. To address this issue, we rounded the values of $\alpha \!=\! \{2.92801, 3.01129, 2.00807, 3.00726, 3.01\}$ to the closest integers. Both the theoretical and simulation results show a strong match for various values of $\alpha$ and $\mu$. Our results provide practical insights for THz system design. The $\alpha$-$\mu$ fading model improves propagation predictions, while hardware impairments ($\kappa_t$, $\kappa_r$) highlight practical limitations. SNR-based SER analysis guides power control, optimizing performance for real-world THz systems.
\begin{table}[t!]
\centering
\caption{Parameter for $\alpha$-$\mu$ distribution}
\begin{tabular}{|c|c|c|c|c|}
\hline
$\alpha$ & $\mu$ & $\hat{Z}$ & $d(m)$ & Ref. \\ \hline
3.45388  & 0.51571 & 6.94184  & 10.27 & \cite{papasotiriou2021experimentally} \\ \hline
3.37619  & 2.77203 & 15.14883 & 3.15  & \cite{papasotiriou2021experimentally} \\ \hline
3.28199  & 1.71725 & 10.74124 & 10.04 & \cite{papasotiriou2021experimentally} \\ \hline
3.01129  & 0.69601 & 8.51768  & 18.06 & \cite{papasotiriou2021experimentally} \\ \hline
2.92801  & 0.61844 & 4.35616  & 26.53 & \cite{papasotiriou2021experimentally} \\ \hline
2.00807  & 1.00254 & 33.37845 & 45.34 & \cite{papasotiriou2021experimentally} \\ \hline
3.00726  & 0.59581 & 65.86256 & 15.36 & \cite{papasotiriou2021experimentally} \\ \hline
3.01  & 1.65 & 41.29 & 6.9 &\cite{papasotiriou2021fading}  \\ \hline
\end{tabular}
\label{Tab1}
\end{table}

\section{Conclusion}
In this work, we developed a performance analysis framework for linear ZF detection in indoor SISO THz channels with $\alpha$-$\mu$ small-scale fading. We derived the SER under both independent and correlated channel-noise conditions, proposing different methods to capture correlation. Incorporating channel-correlated noise improves model accuracy and enables more physically consistent studies of mitigation techniques and network performance. We also introduced a novel Gaussian-over-$\alpha$-$\mu$ distribution. Theoretical results align closely with simulations, confirming the framework’s accuracy and robustness. This approach provides valuable insights for performance analysis in emerging communication systems and can be extended to other models and detection schemes.

\bibliographystyle{IEEEtran}
\bibliography{IEEEabrv,my_bibliography}

\end{document}